\documentclass[letter]{sig-alternate-hotpower13}
\newcommand{\be}{\begin{itemize}} \newcommand{\ee}{\end{itemize}}

\usepackage{epsfig}
\usepackage{epsfig}
\usepackage{amsmath, amssymb}
\usepackage{subfigure}
\usepackage{color}
\usepackage{algorithmic}
\usepackage{algorithm}
\usepackage[table]{xcolor}
\usepackage[center]{caption}
\usepackage{eurosym}
\usepackage[toc,page]{appendix}

\makeatletter
\def\@copyrightspace{\relax}
\makeatother

\begin{document}

\title {Evolution of IEEE 802.11 compatible standards \\and impact on Energy Consumption}

\numberofauthors{1} 
\author{\alignauthor Stratos Keranidis, Giannis Kazdaridis, Nikos Makris, \\Thanasis Korakis, Iordanis Koutsopoulos and Leandros Tassiulas \\
\affaddr{Department of Computer and Communication Engineering, University of Thessaly, Greece} \\ \affaddr{ Centre for Research and Technology Hellas, CERTH, Greece} \\
\email{\{efkerani, iokazdarid, nimakris, korakis, jordan, leandros\}@uth.gr}
} 

\maketitle

\begin{abstract}

Over the last decade, the IEEE 802.11 has emerged as the most popular protocol in the wireless domain.
Since the release of the first standard version, several amendments have been introduced in an effort
to improve its throughput performance, with the most recent one being the IEEE 802.11n extension.
In this work, we present detailed experimentally obtained measurements that evaluate the energy efficiency of the base standard
in comparison with the latest 802.11n version.
Moreover, we investigate the impact of various MAC layer enhancements, both vendor specific and standard compliant ones,
on the energy consumption of wireless transceivers and total nodes as well.
Results obtained under a wide range of settings, indicate that the latest standard enables reduction of energy expenditure, 
by more than 75$\%$, when combined with innovative frame aggregation mechanisms.

\end{abstract}


\vspace{-0.2cm}
\section{Introduction}
\vspace{-0.1cm} 
IEEE 802.11 is currently considered as the default solution for implementing wireless local 
area network communications.
The wide adoption of this standard by vendors of wireless devices offers high interoperability, which in combination with the provided ease of use and low
deployment cost have resulted in its unprecedented market and everyday life penetration.
While the base version of the standard was released in 1997, subsequent amendments have been proposed throughout the years,
among which 802.11b, 802.11a and 802.11g are the most widely accepted ones.
In 2007, the current standard IEEE 802.11-2007 \cite{IEEE802.11-2007} was released and merged several amendments 
with the base version. 

The aforementioned versions of IEEE 802.11 use different PHY layer specifications, but are all based on the same MAC architecture.
The mandatory access scheme that has been specified by the legacy IEEE 802.11 standard is implemented through the distributed coordination function (DCF)
that is based on the carrier sense multiple access with collision avoidance (CSMA/CA) mechanism.
The large PHY and MAC layer overheads that are associated with the DCF process result in a reduction of more than 50$\%$ of the nominal link capacity,
which effect is more pronounced for higher PHY bit rates, as shown in \cite{FF_IVAN}.
Moreover, the work in \cite{80211_LIMITS} has shown that when the frame size is 1024 bytes and the PHY bit rate is 54 Mbps,
the maximum achievable throughput is upper bounded to 50.2 Mbps, for infinitely high PHY bit rate.
As a result, it was confirmed that throughput improvement could only be achieved through MAC layer enhancements that would reduce the impact of PHY and MAC layer overheads.

\vspace{-0.2cm}
\section{Evolution of IEEE 802.11}
\vspace{-0.1cm} 
In an effort to improve throughput performance, vendors of wireless products started integrating innovative techniques into their products, as early as 2003.
Such techniques include the "Atheros Fast Frames" (\emph{FF}) \cite{FF_ARTICLE},
which improves 802.11a/b/g performance, by combining two MAC frames into the payload of a single aggregated frame. 
However, application of vendor-specific techniques resulted in hardware incompatibilities, or at least interfered communications \cite{BROADCOM}.



                            \begin{figure*}[!t]
      \begin{center}
      \subfigure[AR5424]{
      \includegraphics[width=0.6\columnwidth]{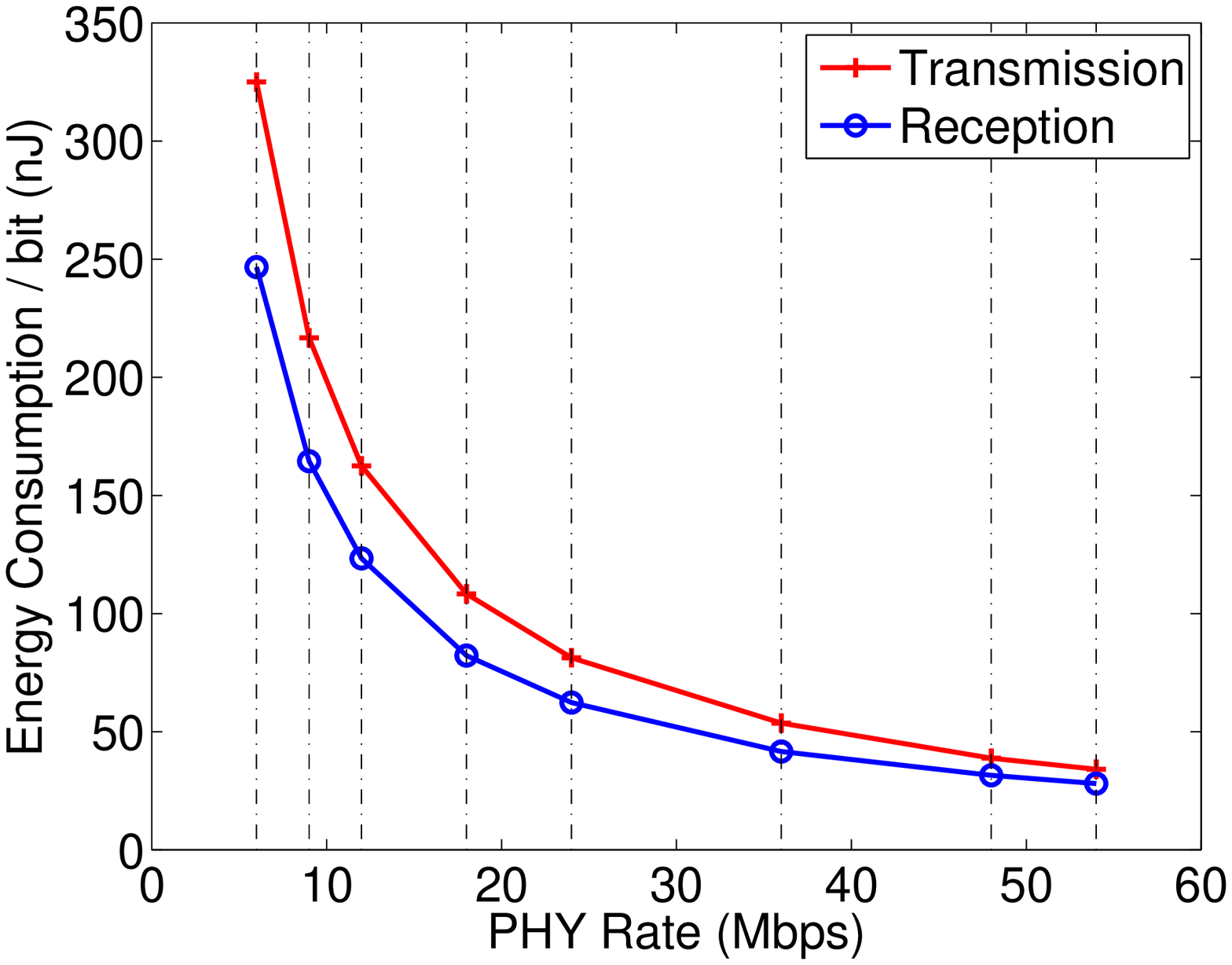}
      \label{fig: Eb_AG}}
      \subfigure[AR9380 Transmission]{
      \includegraphics[width=0.6\columnwidth]{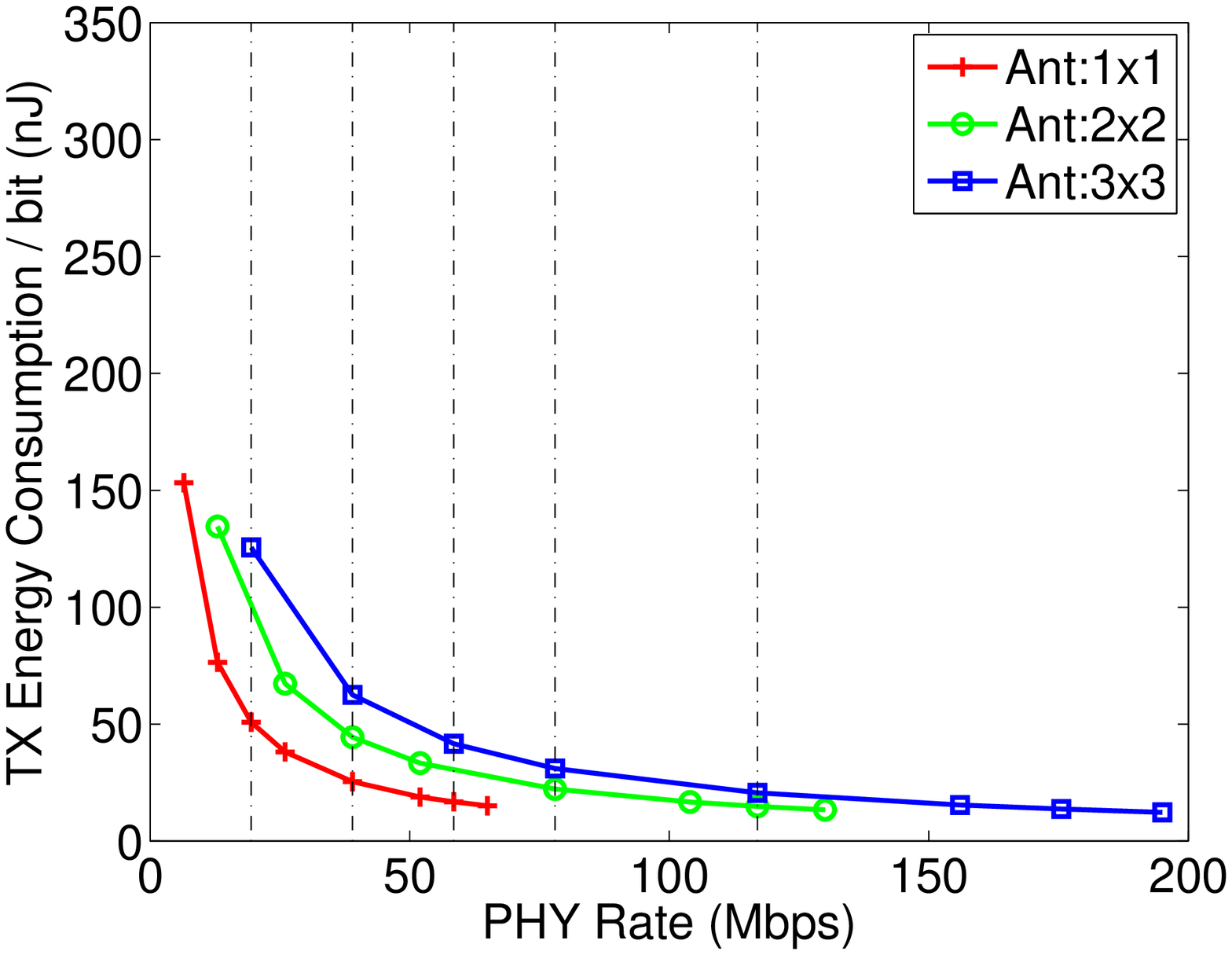}
      \label{fig: EB_MIMO_TX}}
      \subfigure[AR9380 Reception]{
      \includegraphics[width=0.6\columnwidth]{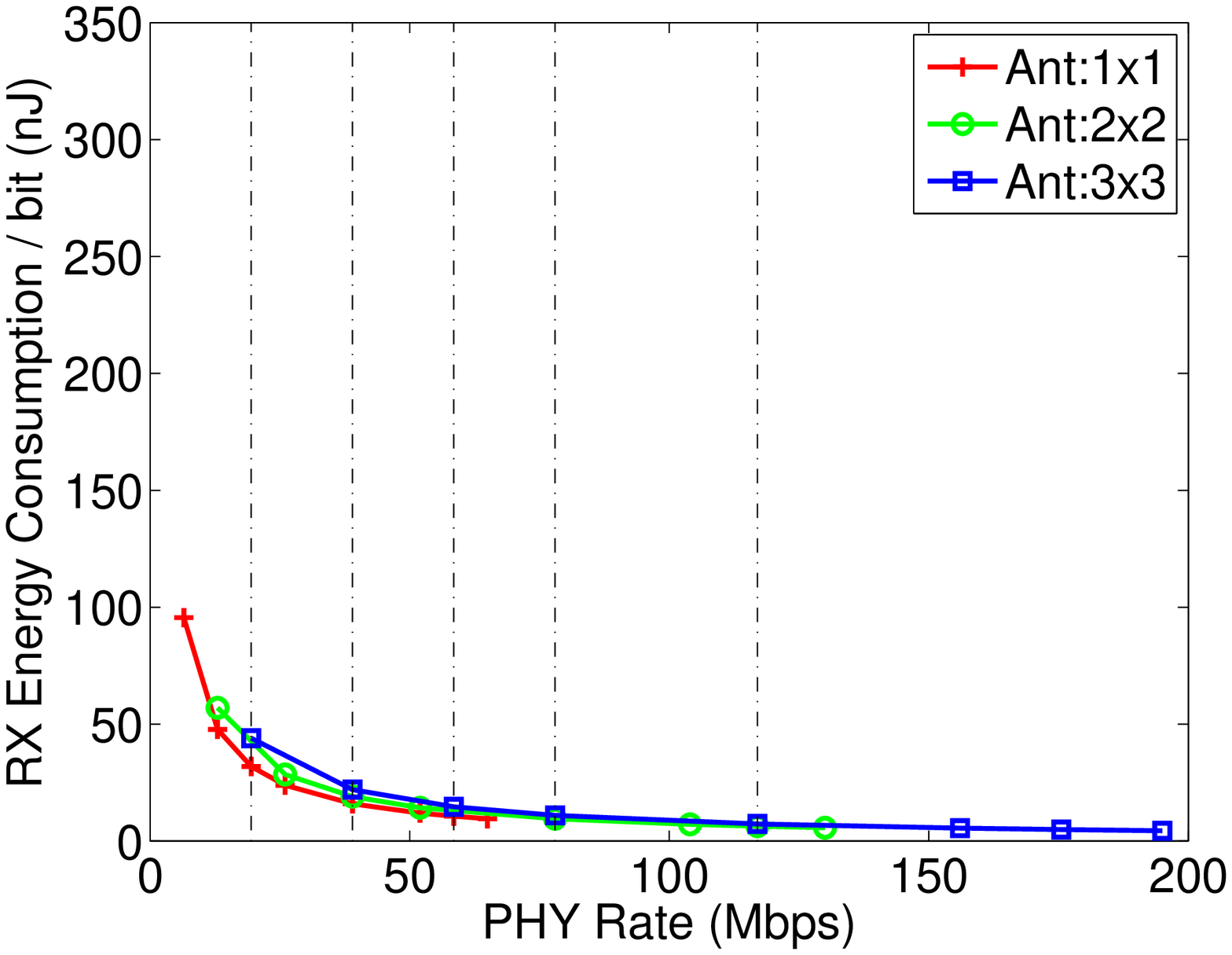}
      \label{fig: EB_MIMO_RX}}
                   \vspace{-0.5cm}
          \caption{Energy consumption/bit at NIC level across available PHY bit rate configurations}
           \vspace{-0.8cm}
      \end{center}
      \end{figure*}  
      
In the same direction, the IEEE 802.11 standard working group started in 2003 to develop the IEEE 802.11n high-throughput (HT)
extension of the base standard that was finally published in 2009.
802.11n offers both PHY and MAC layer enhancements over legacy 802.11 systems.
The most important improvement of the 802.11n on the PHY layer is the ability to
combine multiple antenna elements to achieve higher PHY bit rates and increased link reliability
through the exploitation of multi-stream transmissions and antenna diversity \cite{MIMO}.
Another significant feature is the channel bonding, which increases the channel
bandwidth from 20 MHz to 40 MHz and thus doubles the theoretical capacity limits.
Application of the aforementioned enhancements in combination with several other PHY-layer features
are able to deliver the remarkably increased PHY bit rate of 600 Mbps, resulting in an improvement of more than 10x compared to the legacy 802.11a/g systems.       
In order to increase medium utilisation and exploit from the increased PHY bit rates,
two different types of frame aggregation are provided, namely A-MSDU
and A-MPDU aggregation.
The former combines multiple higher layer packets into a single MAC layer frame with maximum size of 7935 bytes,
while the later combines multiple MAC layer frames to 
form an aggregated frame that cannot exceed the 65.536 bytes. 
In general, A-MPDU aggregation outperforms A-MSDU, 
whose performance is considerably degraded under low quality channel conditions and high PHY rates, as it is shown in \cite{AMPDU_PERFORMANCE}.
Both frame aggregation mechanisms are enhanced by a block acknowledgment mechanism, which further reduces overhead.
In this work, we experimentally investigate how the evolution of the 802.11 standard has impacted the energy consumption 
of wireless transceivers and total nodes, under a wide range of settings.

              \begin{figure*}[!t]
            \vspace{-0.5cm} 
                        \centering
            \subfigure[AR5424]{
      \centering
      \includegraphics[width=0.6\columnwidth]{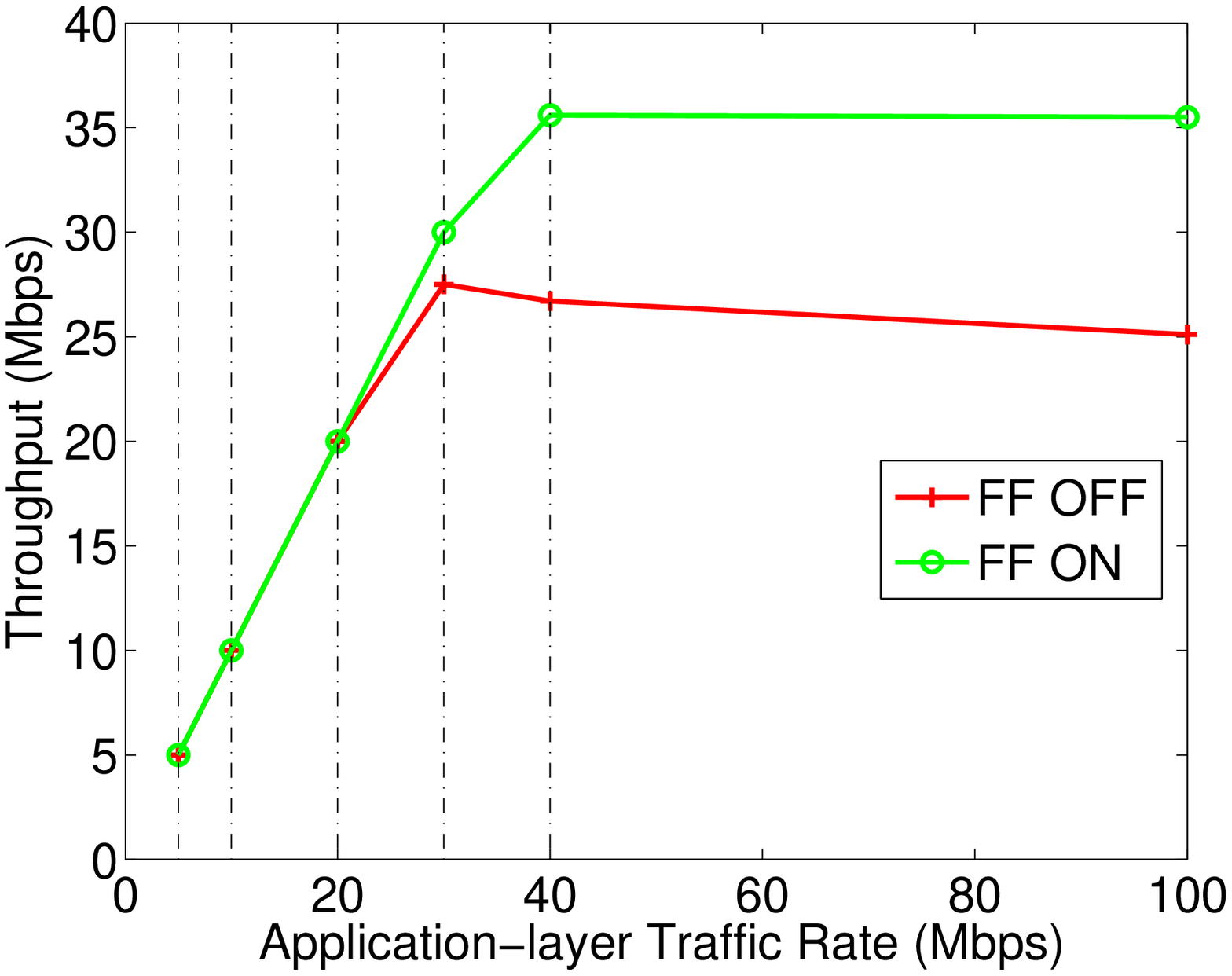}
      \label{fig: THR_AG}}
            \hspace{0.3in}
      \centering
      \subfigure[AR9380]{
      \centering
      \includegraphics[width=0.6\columnwidth]{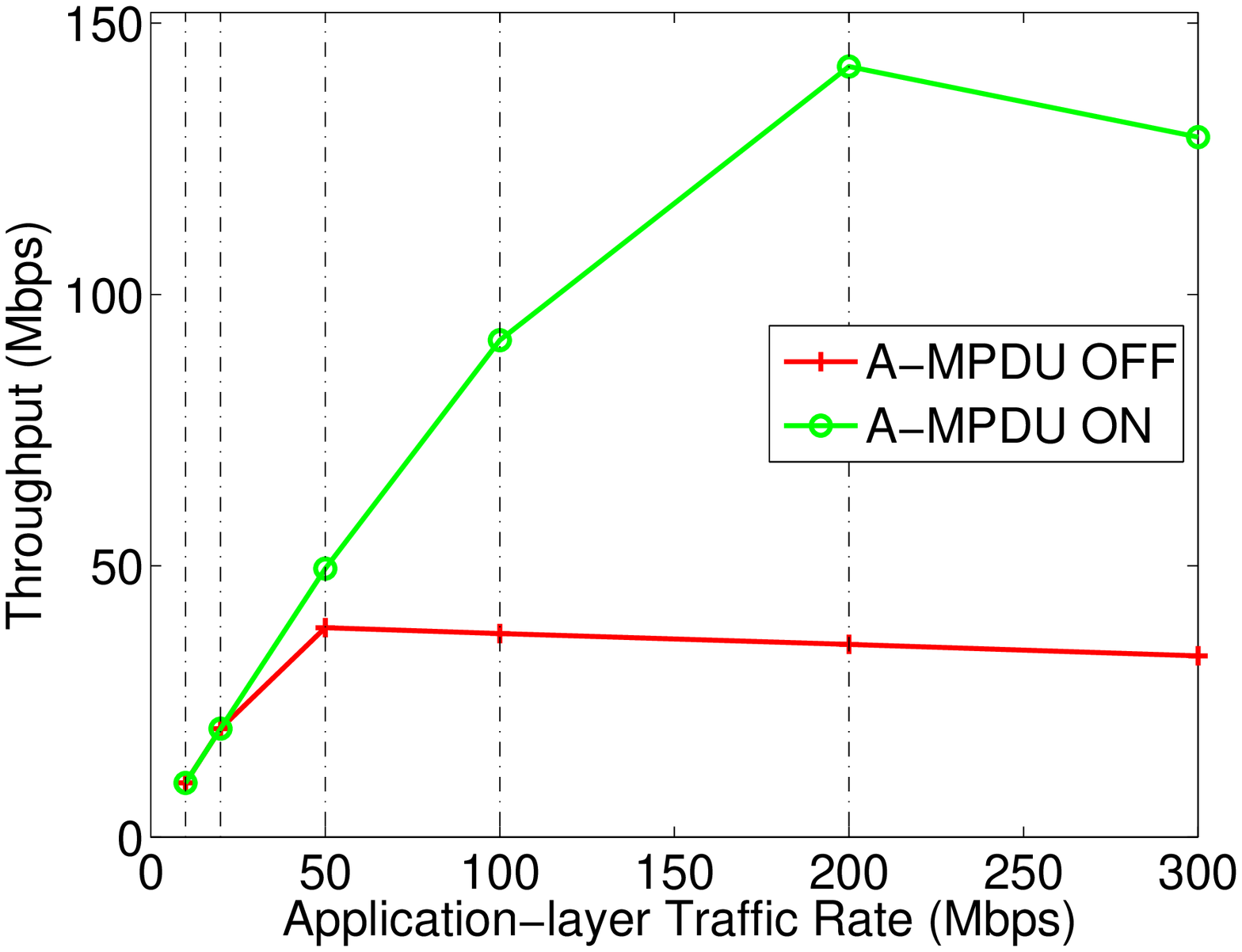}
      \label{fig: THR_MIMO}}
                   \vspace{-0.5cm}
          \caption{Throughput Performance per NIC across varying Application-Layer Traffic load}
           \vspace{-0.5cm}
      \end{figure*}  
       
                            \begin{figure*}[!t]
      \begin{center}
                        \vspace{-0.1cm}
      \subfigure[AR5424]{
      \includegraphics[width=0.6\columnwidth]{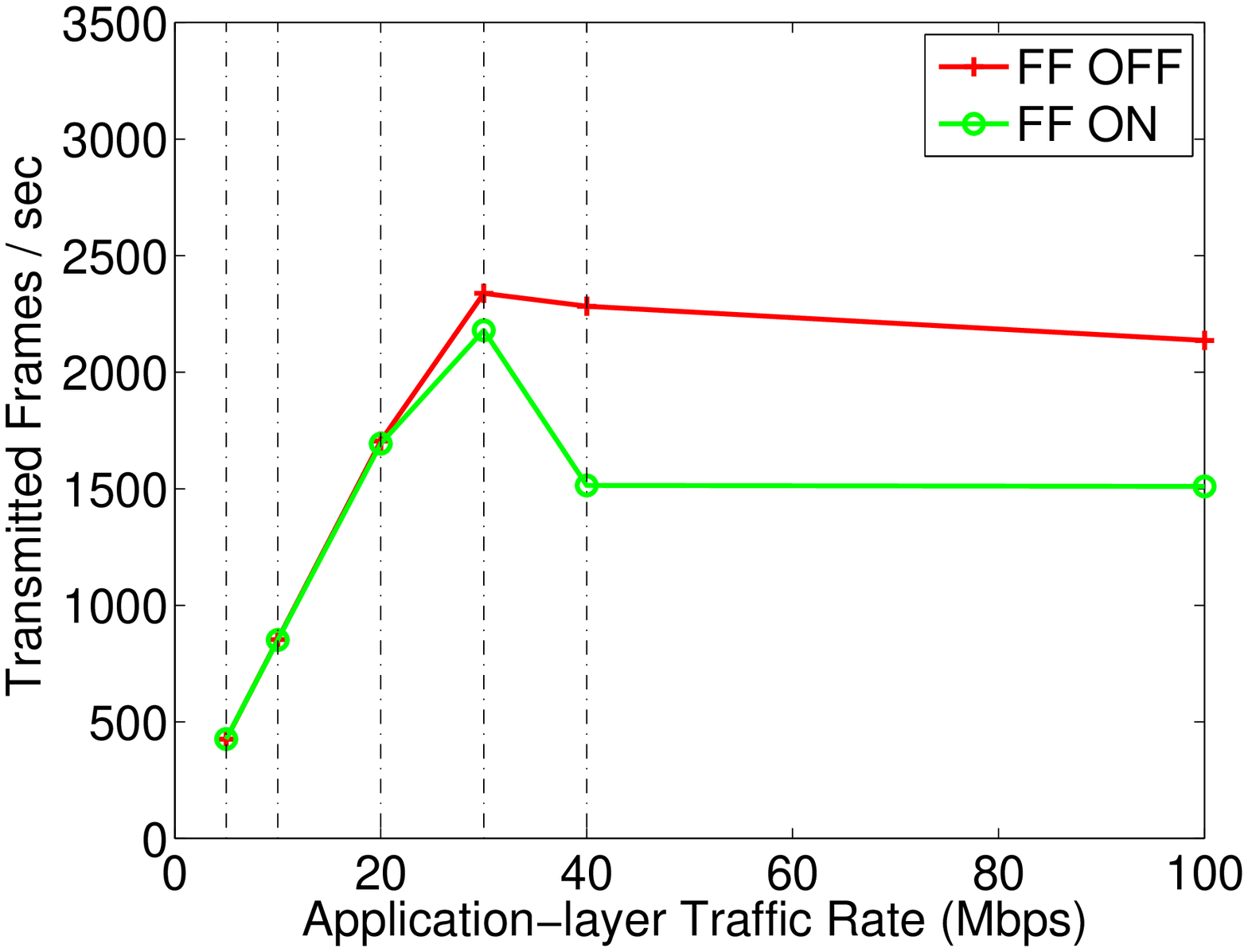}
      \label{fig: FRA_AG}}
      \hspace{0.3in}
      \subfigure[AR9380]{
      \includegraphics[width=0.6\columnwidth]{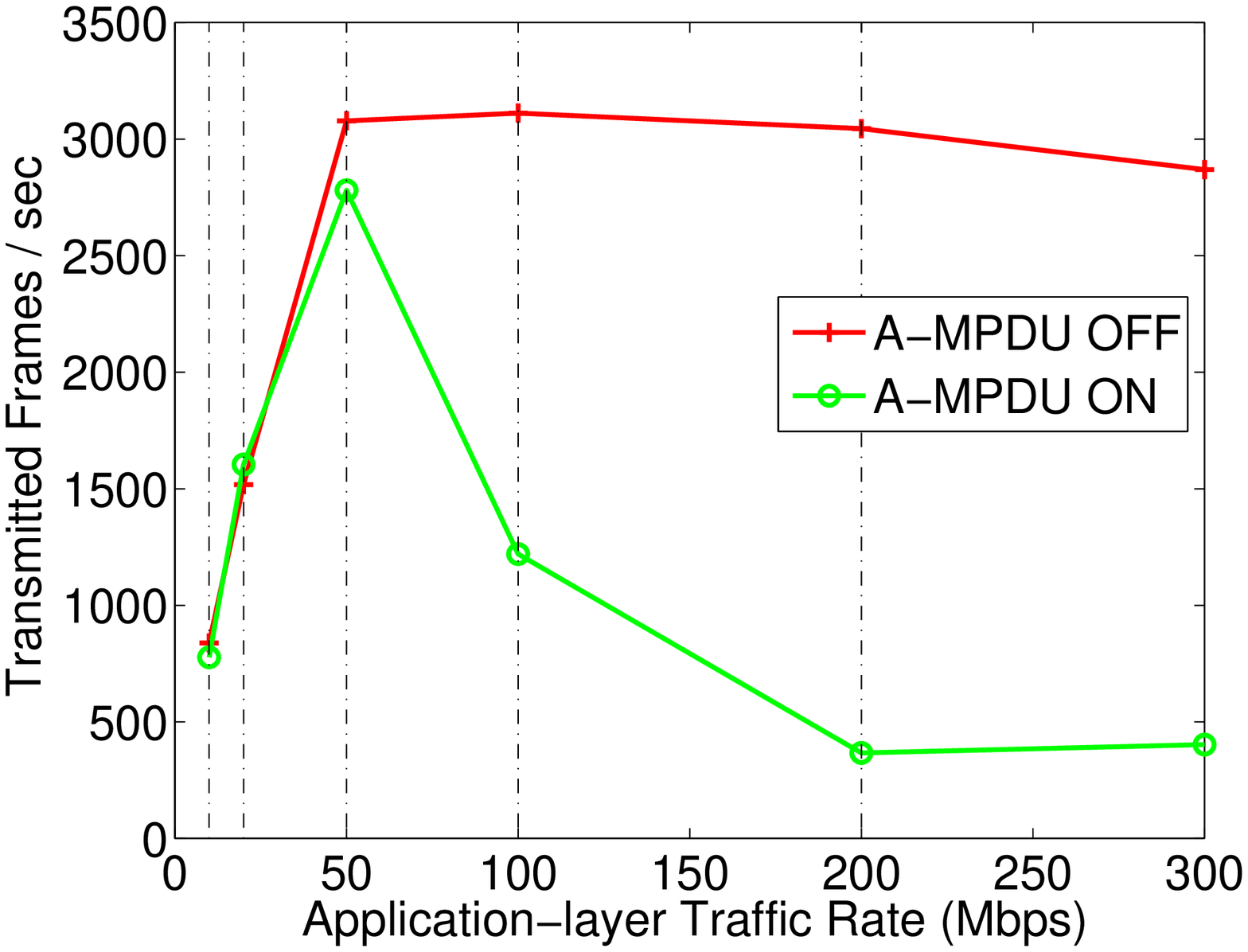}
      \label{fig: FRA_MIMO}}
                   \vspace{-0.5cm}
          \caption{Transmitted MAC-layer Frames / sec per NIC across varying Application-Layer Traffic load}
           \vspace{-0.8cm}
      \end{center}
      \end{figure*}

\vspace{-0.2cm} 
\section{Experimental Evaluation}
\vspace{-0.1cm} 
In this section, we conduct several experiments that aim at comparing the performance of 802.11a/g and 802.11n standards.
For this purpose, we chose two commercial wireless NIC that are representative of the state-of-the-art of each standard.
More specifically, we use the Atheros AR5424 and AR9380 chipsets that implement the 802.11a/g and 802.11n protocols
and are configured through the use of the \emph{Mad-WiFi} \cite{MADWIFI} and \emph{ath9k} \cite{ATH9K} open source drivers accordingly.
We use the same experimental setup while evaluating performance of the two protocols,
which consists of a single communicating pair of nodes that both feature the Intel Atom processor D525 (1.8 GHz) and operate in infrastructure mode.
The wireless nodes are closely spaced within an office environment and operate on the RF-isolated channel 36 of the 5 GHz band,
so that high quality channel conditions are constantly guaranteed.
In order to provide for a proper comparison setup between the two standards, 
we fix the same channel bandwidth of 20 MHz and OFDM guard interval of 0.8 $\mu$s for both chipsets.
Under these settings AR9380 supports the maximum PHY bit rate values of 65 Mbps, 130 Mbps and 195 Mbps
for single, double and triple spatial stream configurations accordingly, while AR5424 supports PHY bit rate values between 6 Mbps and 54 Mbps.

\vspace{-0.2cm} 
\subsection{Energy Consumption Characterisation}
\vspace{-0.1cm} 
Energy consumption measurements are executed on the station node, where current shunt resistors of 0.1$\Omega$ and 0.01$\Omega$ have been placed
in series with the power supply of the NIC and the Atom-based node accordingly.
We use the prototype NITOS ACM card to accurately measure the voltage drop across the resistors at the high sampling rate of 63 KHz
and thus we are able to jointly evaluate instantaneous power consumption of the NIC and the total Atom-based node as well.
More details about the developed card and the followed measurement procedure can be found in our previous work \cite{TR}.
We start by characterising the instantaneous power consumption of the two NICs across various operational modes and present results in Table 1.

                 \vspace{-0.1cm}
      \begin{table}[!h] 
        \begin{center}
      \scalebox{0.9} {
          \begin{tabular}{| c | c | c | c | c |}
          \hline
         \cellcolor[gray]{0.7}  Chipset  & \cellcolor[gray]{0.7} AR5424  
         &  \multicolumn{3}{c}{\cellcolor[gray]{0.7} AR9380} \vline\\ \hline  
          Antennas & 1x1 & 1x1 & 2x2 & 3x3\\ \hline
          \cellcolor[gray]{0.7} Mode &  \multicolumn{4}{c}{\cellcolor[gray]{0.7} Power Consumption (Watts)}  \vline \\ \hline 
           Sleep & -  &  \multicolumn{3}{c}{0.12} \vline \\ \hline
           Idle & 1.47 & 0.49 & 0.56 & 0.69\\ \hline
           Receive & 1.52 & 0.62 & 0.74 & 0.85 \\ \hline
           Transmit & 1.97 & 0.98 & 1.75 & 2.45\\ \hline
          \end{tabular}
      }
      \vspace{-0.5cm}
       \end{center}
        \caption{Power consumption of AR5424 and AR9380 NICs across different operational modes}
                   \vspace{-0.4cm}
        \label{table: Selfish Strategy Performance}
      \end{table}

Regarding the sleep mode power consumption of the AR5424 NIC, we remark that we were not able to activate it through the \emph{Mad-WiFi} driver.
On the other hand, we configured \emph{Power Saving Mode (PSM)} mode for the AR9380 NIC,
which set the NIC in low-power state and disabled most of its circuitry.
Regarding the rest modes, we activated only the number of antennas required in each setting and observed that the 802.11n chipset 
consumes significantly less power in the idle and reception modes,
while consumption increases above the levels of the 802.11a/g NIC only in the case that the MIMO 3x3 transmissions are executed.

We also conducted more detailed measurements to calculate the Energy Consumption per bit ($E_{B}$),
under the available PHY bit rates that are supported by each standard, across  frame transmission and reception operations. 
Fig. \ref{fig: Eb_AG} illustrates the obtained $E_{B}$ across the the available 802.11a/g PHY bit rates,
while Fig. \ref{fig: EB_MIMO_TX} and Fig. \ref{fig: EB_MIMO_RX} plot $E_{B}$ across the available 802.11n PHY bit rates, under frame transmission and reception accordingly.
We clearly observe that the remarkably higher rates of 802.11n protocol, in combination with the lower (in most cases) NIC power consumption
result in significantly lower $E_{B}$ values, in comparison with the performance of the earlier 802.11a/g protocol.

                                \begin{figure*}[!t]
                        \centering
            \subfigure[NIC]{
                             \hspace{-0.2cm}
      \centering
      \includegraphics[width=0.7\columnwidth]{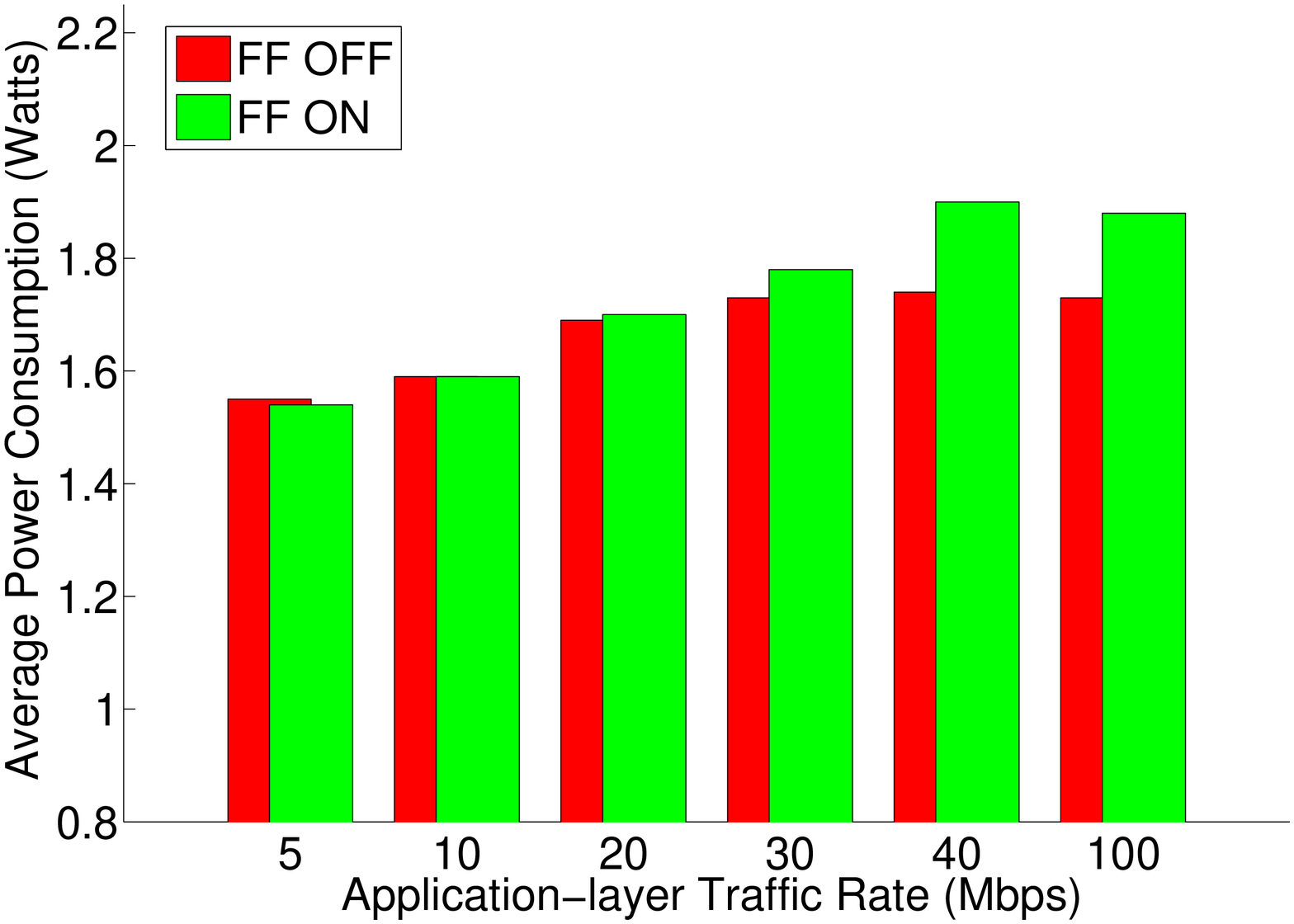}
      \label{fig: FF_NIC}}
                 \hspace{-0.1cm}
      \centering
      \subfigure[Atom Node]{
      \centering
      \includegraphics[width=0.7\columnwidth]{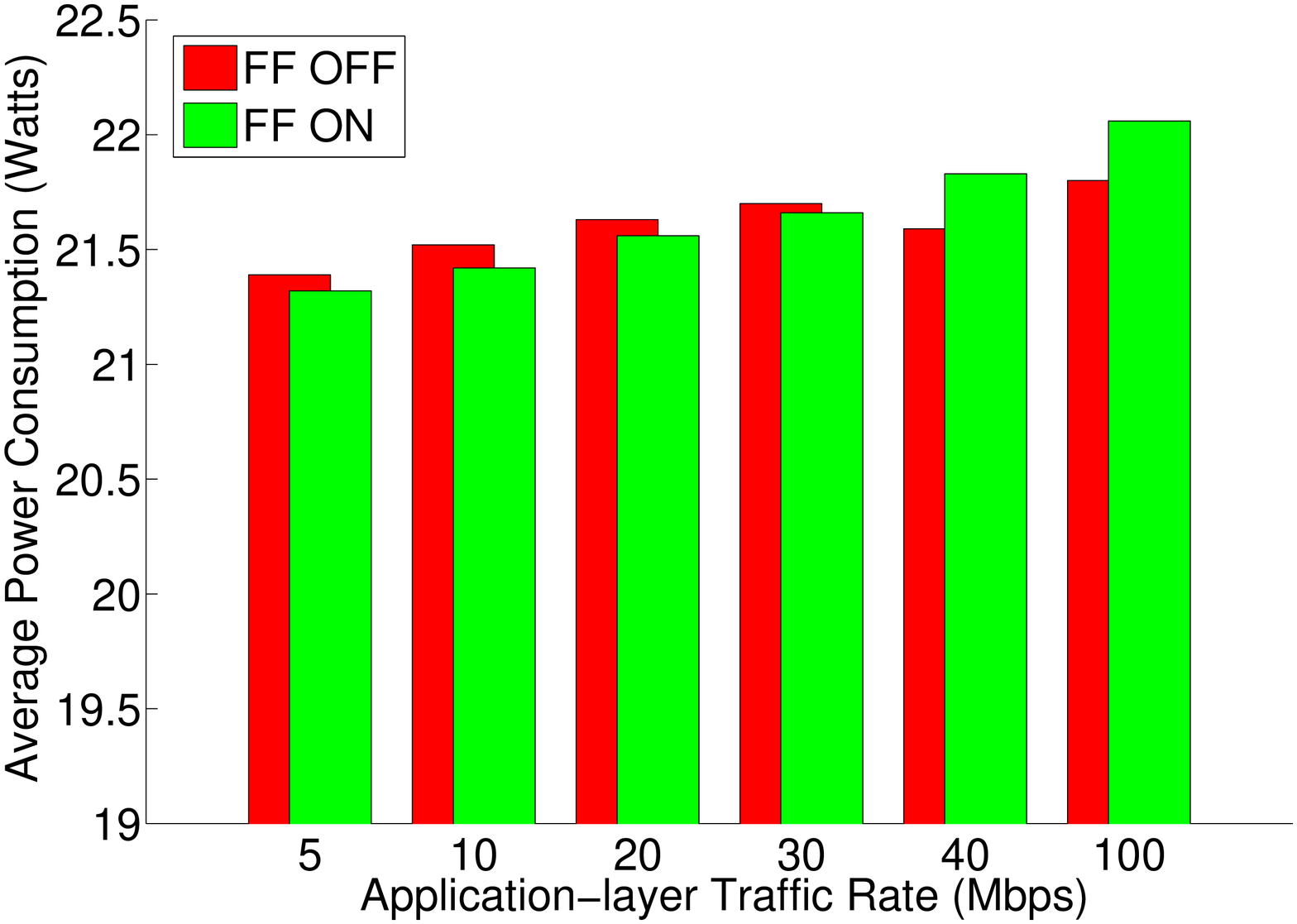}
      \label{fig: FF_ATOM}}
                        \centering
            \subfigure[Atom Energy consumption/bit]{
                  \centering
      \includegraphics[width=0.59\columnwidth]{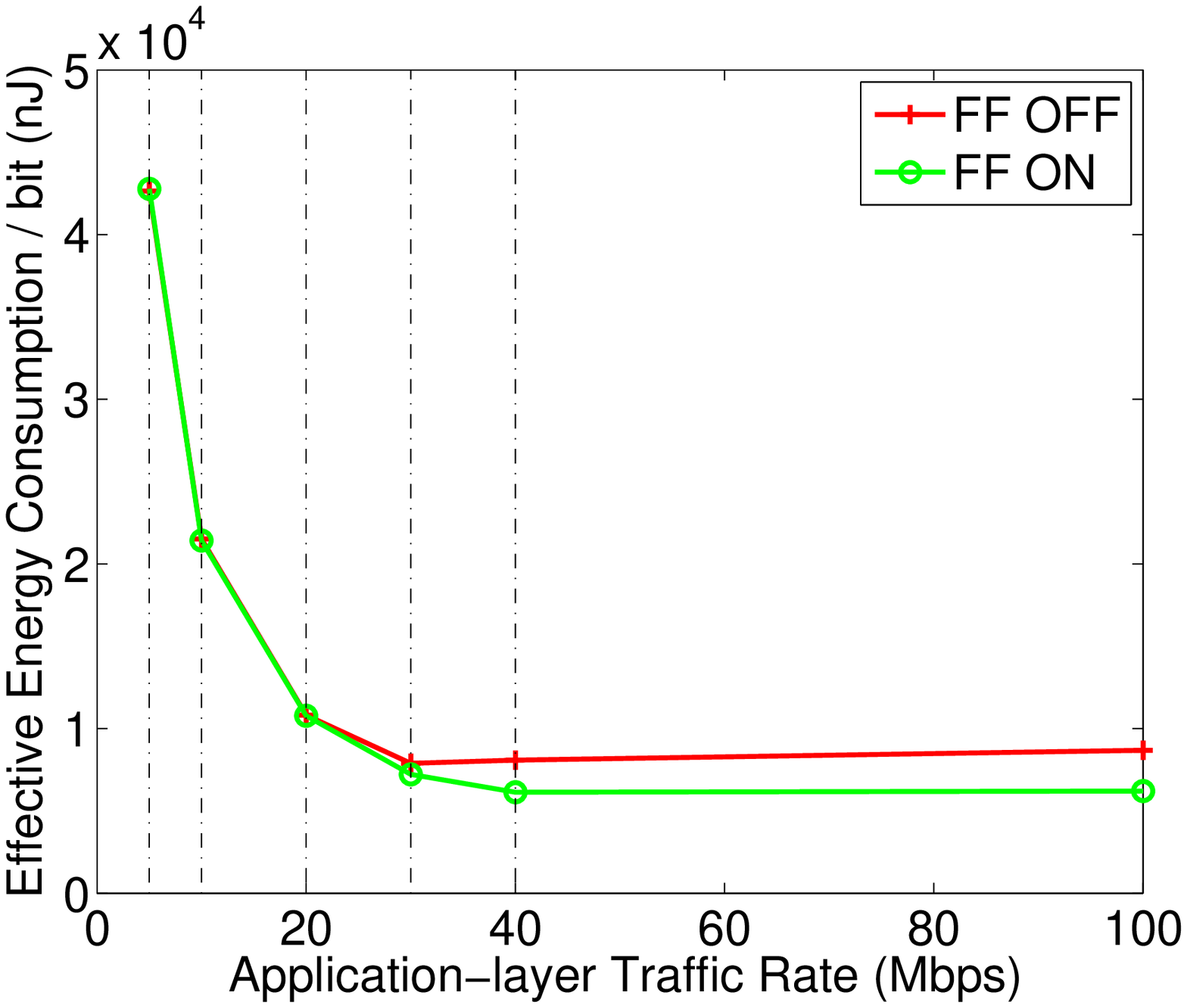}
      \label{fig: EB_AG}}
                   \vspace{-0.5cm}
          \caption{Energy efficiency characterisation of AR5424 setup across varying Application-Layer Traffic load}
           \vspace{-0.4cm}
      \end{figure*}  

                 \begin{figure*}[!t]
            \vspace{-0.1cm}
                        \centering
            \subfigure[NIC]{
                                         \hspace{-0.2cm}
      \centering
      \includegraphics[width=0.7\columnwidth]{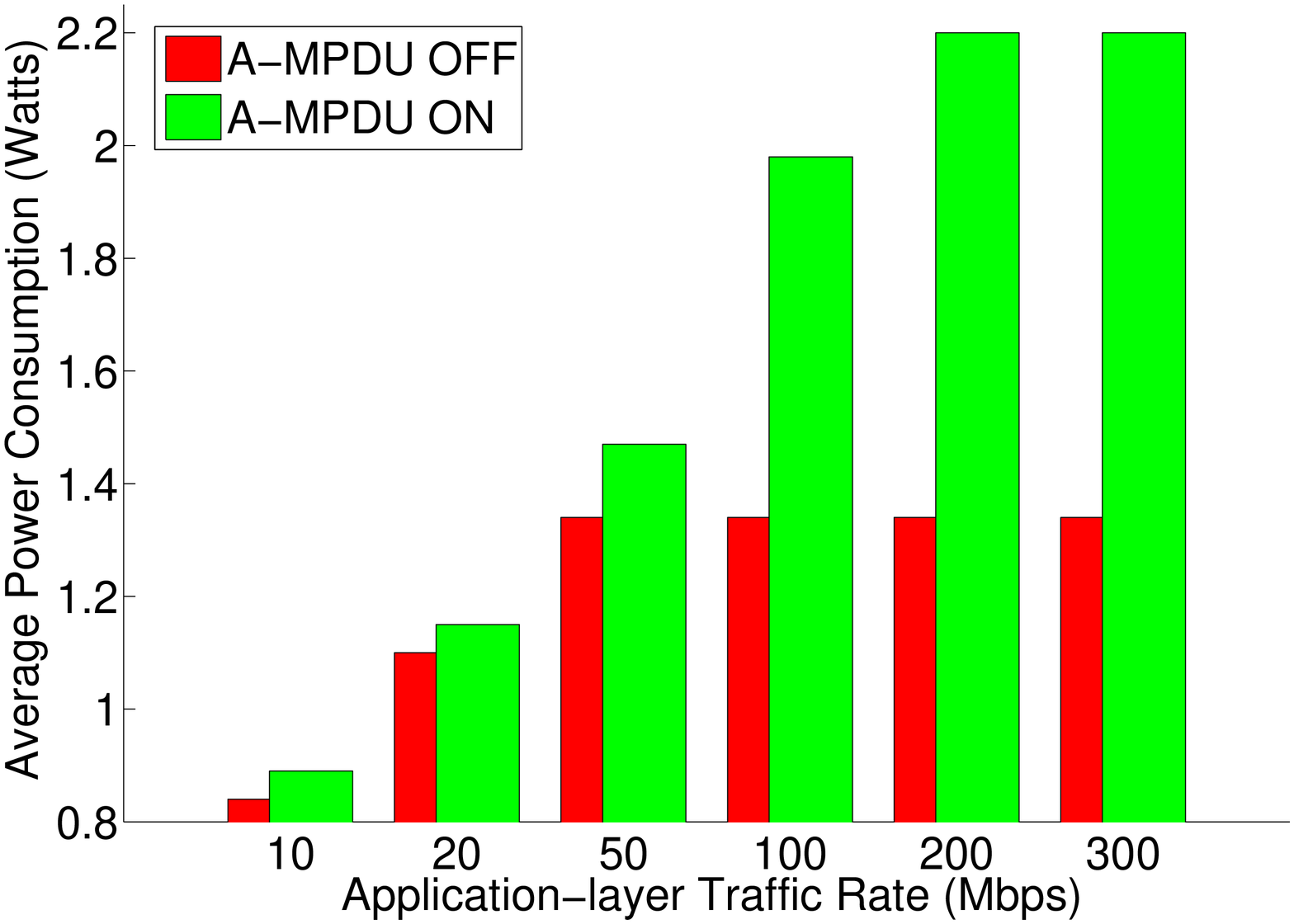}
                             \hspace{-0.1cm}
      \label{fig: AGG_NIC}}
      \centering
      \subfigure[Atom Node]{
      \centering
      \includegraphics[width=0.7\columnwidth]{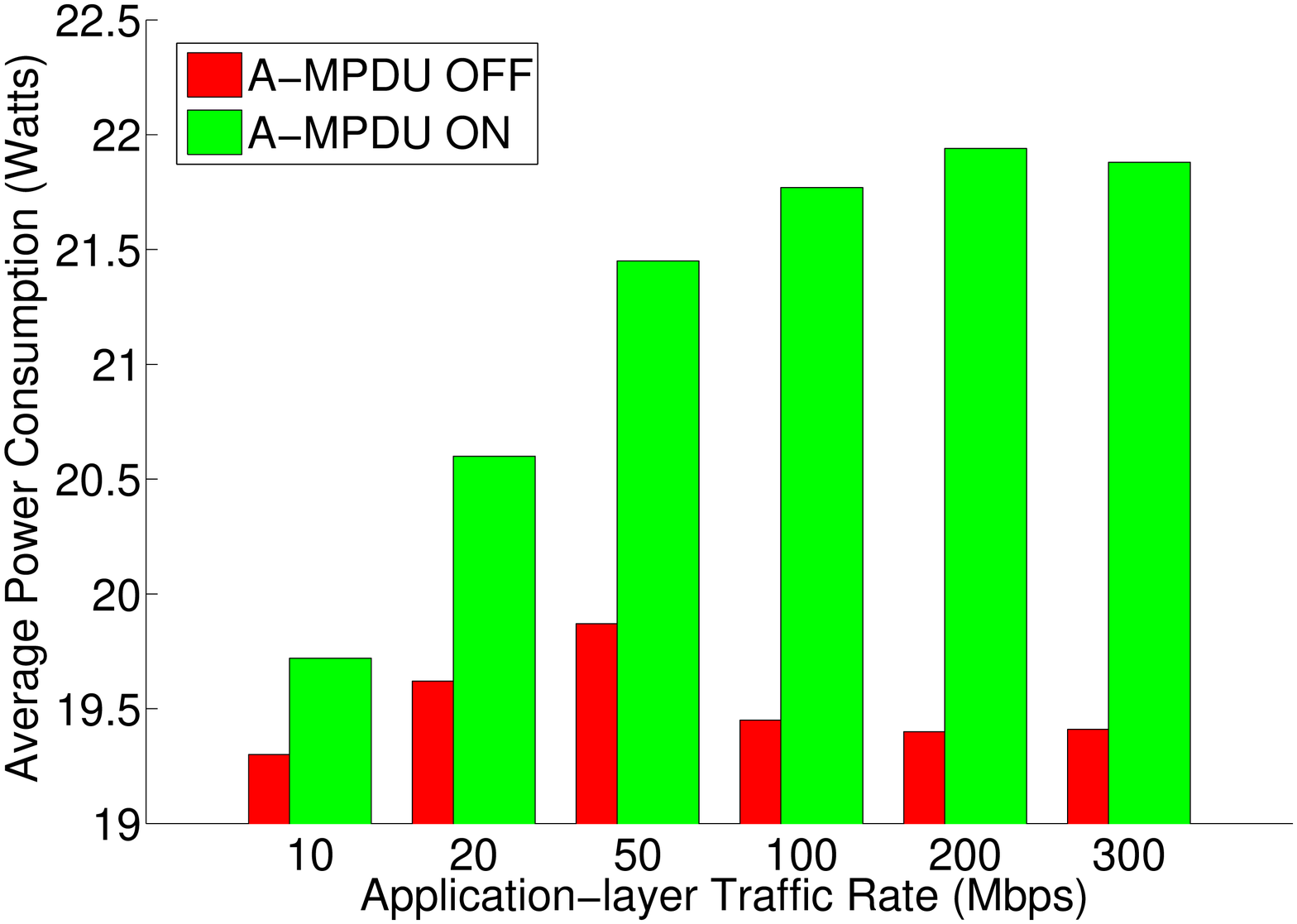}

      \label{fig: AGG_ATOM}}
                        \centering
            \subfigure[Atom Energy consumption/bit]{
                  \centering
      \includegraphics[width=0.59\columnwidth]{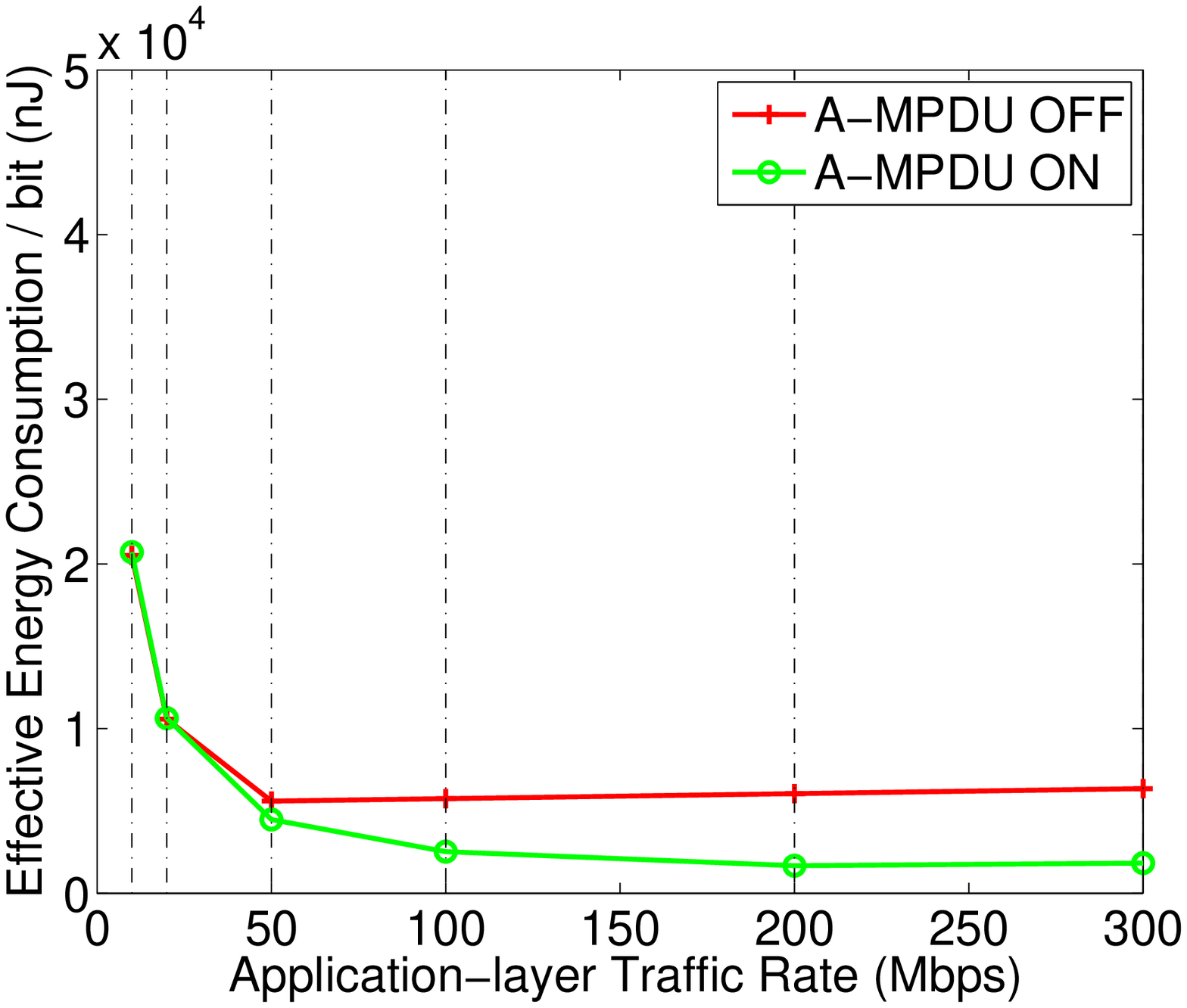}
      \label{fig: EB_MIMO}}
                   \vspace{-0.5cm}
          \caption{Energy efficiency characterisation of AR9380 setup across varying Application-Layer Traffic load}
           \vspace{-0.5cm}
      \end{figure*}  

\vspace{-0.2cm}
\subsection{Realistic Throughput Experiments}
\vspace{-0.1cm}
In this set of experiments, we take a step further from characterising energy consumption under fixed modes of operation
and compare performance of the two standards under realistic throughput experiments. 
We manually configure the maximum available PHY bit rates for each standard
and guarantee that these rates can be supported by the prevailing channel conditions,
by constantly monitoring the achievable frame delivery ratio and assuring that it never drops below 95$\%$ in all the conducted experiments.
We start by measuring throughput performance of each protocol without enabling any form of aggregation,
while we next repeat the same experiments by explicitly enabling the \emph{FF} and A-MPDU aggregation mechanisms that are supported by the 
AR5424 and AR9380 NICs, through proper modification of the corresponding driver.
In each experiment, we also monitor the power consumption at both the NIC and the total node level,
in order to assess the impact of the various configured settings on energy consumption.

\vspace{-0.2cm}
\subsubsection{Varying Application-Layer Traffic load}
\vspace{-0.1cm}

Fig. \ref{fig: THR_AG} and Fig. \ref{fig: THR_MIMO} plot the monitored throughput for the 802.11a/g and 802.11n compatible chipsets
across varying Application-layer traffic loads. 
In Fig. \ref{fig: THR_AG}, we observe that below channel saturation, throughput performance is similar, while maximum achievable throughput differentiates
between the two cases as much as 30$\%$ as the load approaches the 40 Mbps value.
Based on detailed study of the \emph{Mad-WiFi} driver, we conclude that \emph{FF} is only activated when the driver detects that
the channel is approaching saturation through inspection of the transmission queue levels.
We also verified our findings by monitoring the number of MAC-layer frames that are being transmitted under each different load,
which results are illustrated in Fig. \ref{fig: FRA_AG}.
The same experiment was repeated using the 802.11n compatible AR9380 chipset as well and similar conclusions were reached.
In the case that A-MPDU aggregation is disabled, channel reaches the saturation point as soon as traffic load equals 50 Mbps,
while in the A-MPDU enabled case, saturation is only reached at the traffic load of 200 Mbps.
In this case, throughput increases due to the activation of A-MPDU aggregation by a factor of 4x.
Our findings clearly verify that MAC layer improvements need to be applied, in order to exploit from the increased PHY bit rates that 802.11n offers.

Fig. \ref{fig: FF_NIC} and Fig. \ref{fig: FF_ATOM} plot average power consumption of the AR5424 and the total Atom node across the various configured traffic load values.
In Fig. \ref{fig: FF_NIC}, we observe that the NIC consumes similar power in cases that the \emph{FF} mechanism is not yet activated,
while as soon as traffic load reaches the 30 Mbps value, \emph{FF} is activated and average power consumption is reduced,
due to the decreased frequency of frame transmissions. 
Under higher load conditions, AR5424 consumes more power on average as it operates in transmit mode for longer duration. 
When considering the power consumption of the total Atom node, we noticed that even under low traffic loads, 
\emph{FF} activation lowers consumption, although it is not yet applied.
We identified that the detected anomaly results due to the different frame handling approach that is followed by the driver across the two different cases,
and results in considerably more efficient function calls in the case that the \emph{FF} option is enabled. 
Our findings are summarised in the Effective $E_{B}$ representation in Fig. \ref{fig: EB_AG}, which characterizes the total node power consumption as a function of the resulting throughput. The obtained results clearly show that \emph{FF} is able to reduce energy expenditure up to 25$\%$.

              \begin{figure*}[!t]
      \begin{center}
            \subfigure[Throughput]{
      \includegraphics[width=0.6\columnwidth]{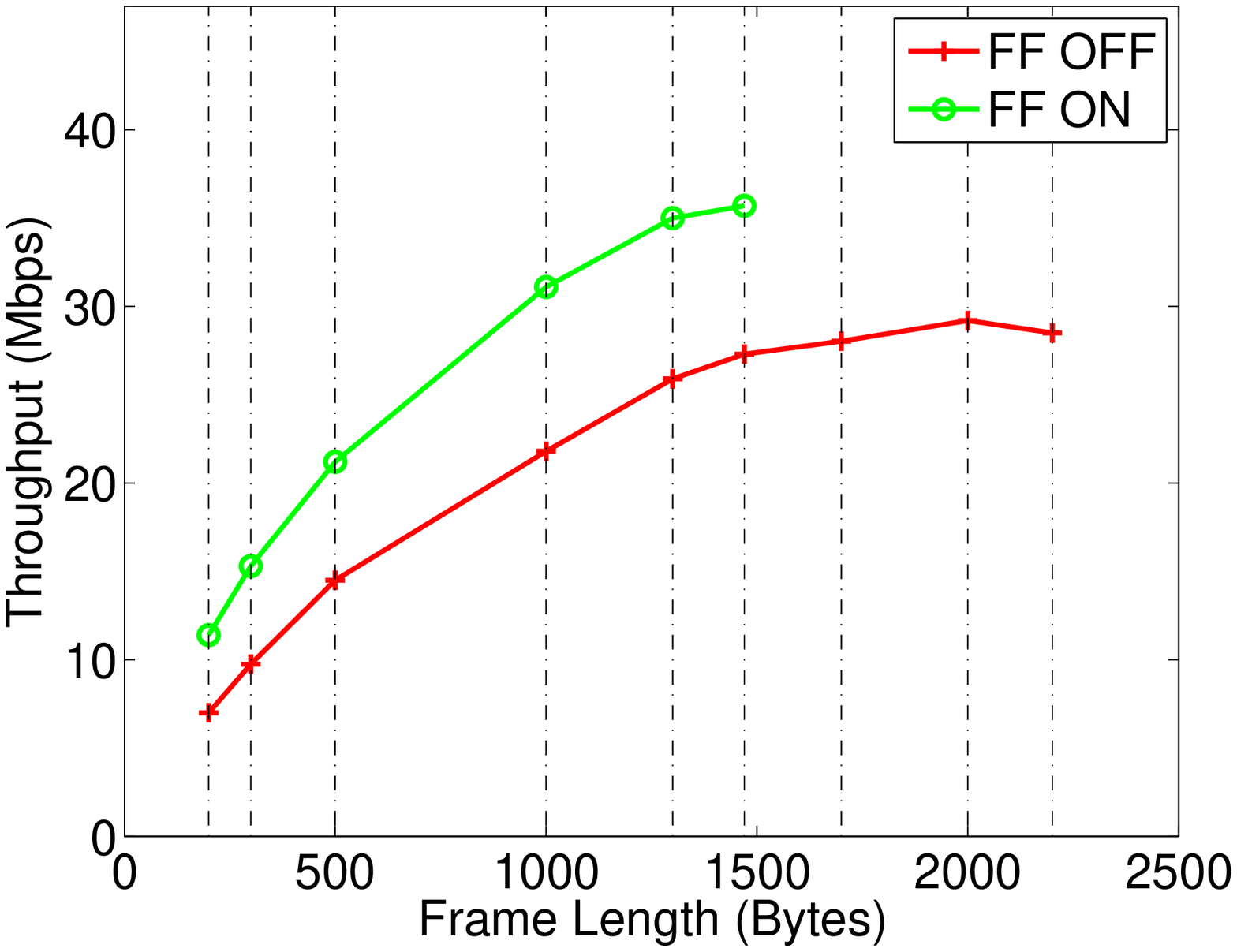}
      \label{fig: THR_LEN_AG}}
      \hspace{0.21in}
      \subfigure[Transmitted Frames / sec]{
      \includegraphics[width=0.6\columnwidth]{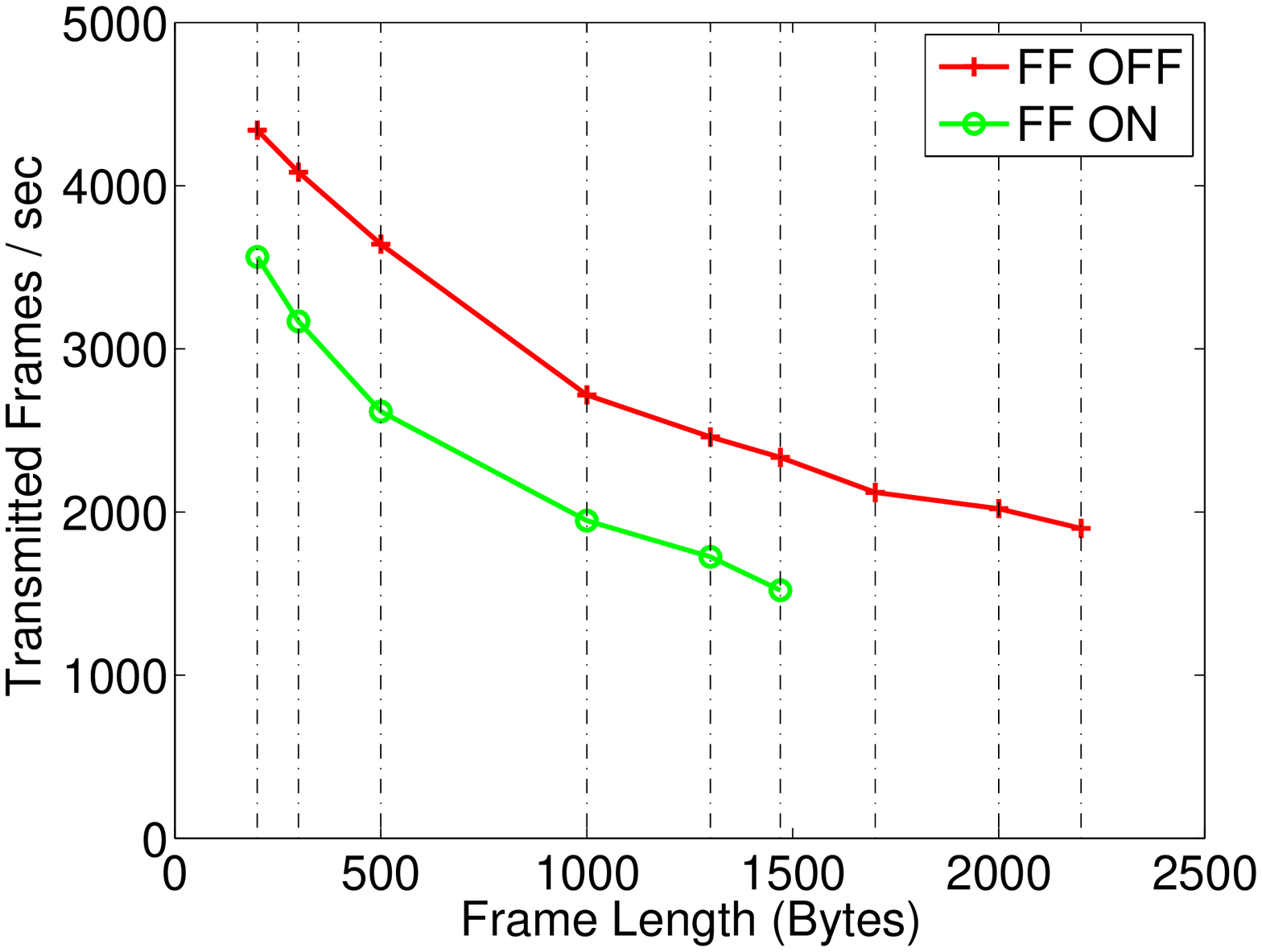}
      \label{fig: FRM_LEN_AG}}
                \hspace{0.22in}             
            \subfigure[Atom Energy consumption/bit]{
      \includegraphics[width=0.6\columnwidth]{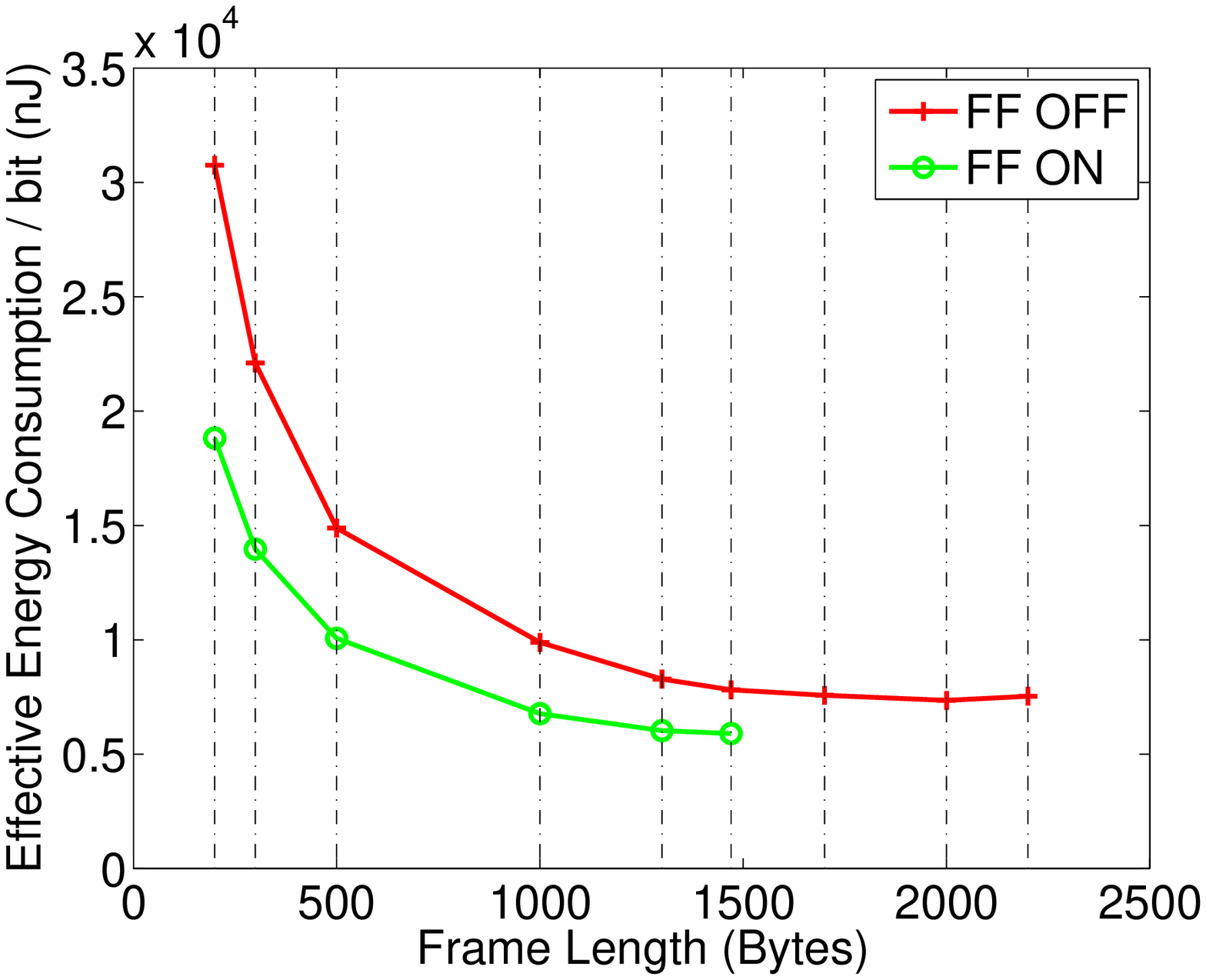}
      \label{fig: EB_LEN_AG}}
             \vspace{-0.4cm}
          \caption{Performance of AR5424 NIC across varying Frame Length values}
           \vspace{-0.7cm}
      \end{center}
      \end{figure*}  

              \begin{figure*}[!t]
                        \vspace{-0.2cm}
                        \centering
            \subfigure[Throughput]{
      \centering
      \includegraphics[width=0.6\columnwidth]{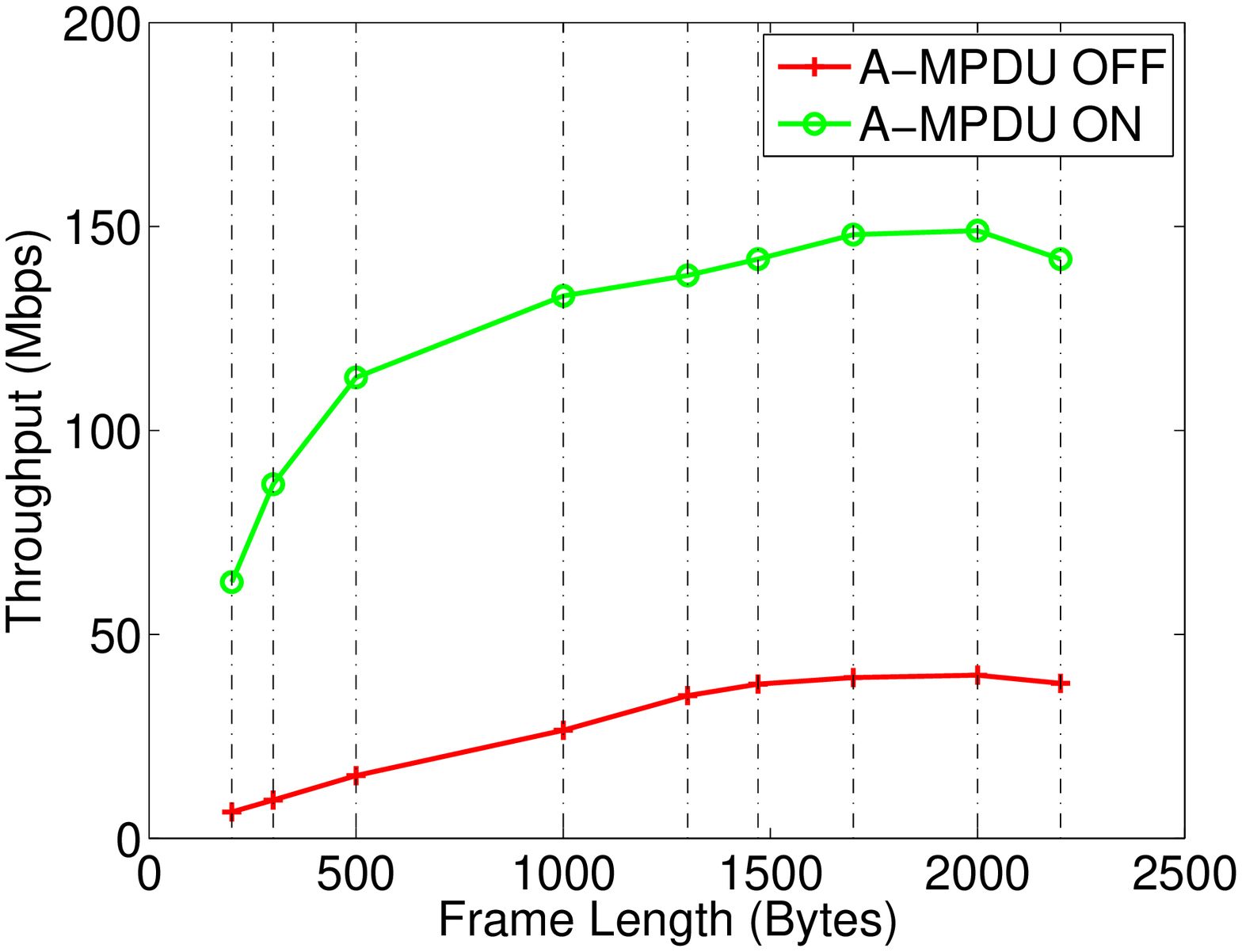}
      \label{fig: THR_LEN_MIMO}}
      \hspace{0.21in}
      \centering
      \subfigure[Transmitted Frames/sec]{
      \centering
      \includegraphics[width=0.6\columnwidth]{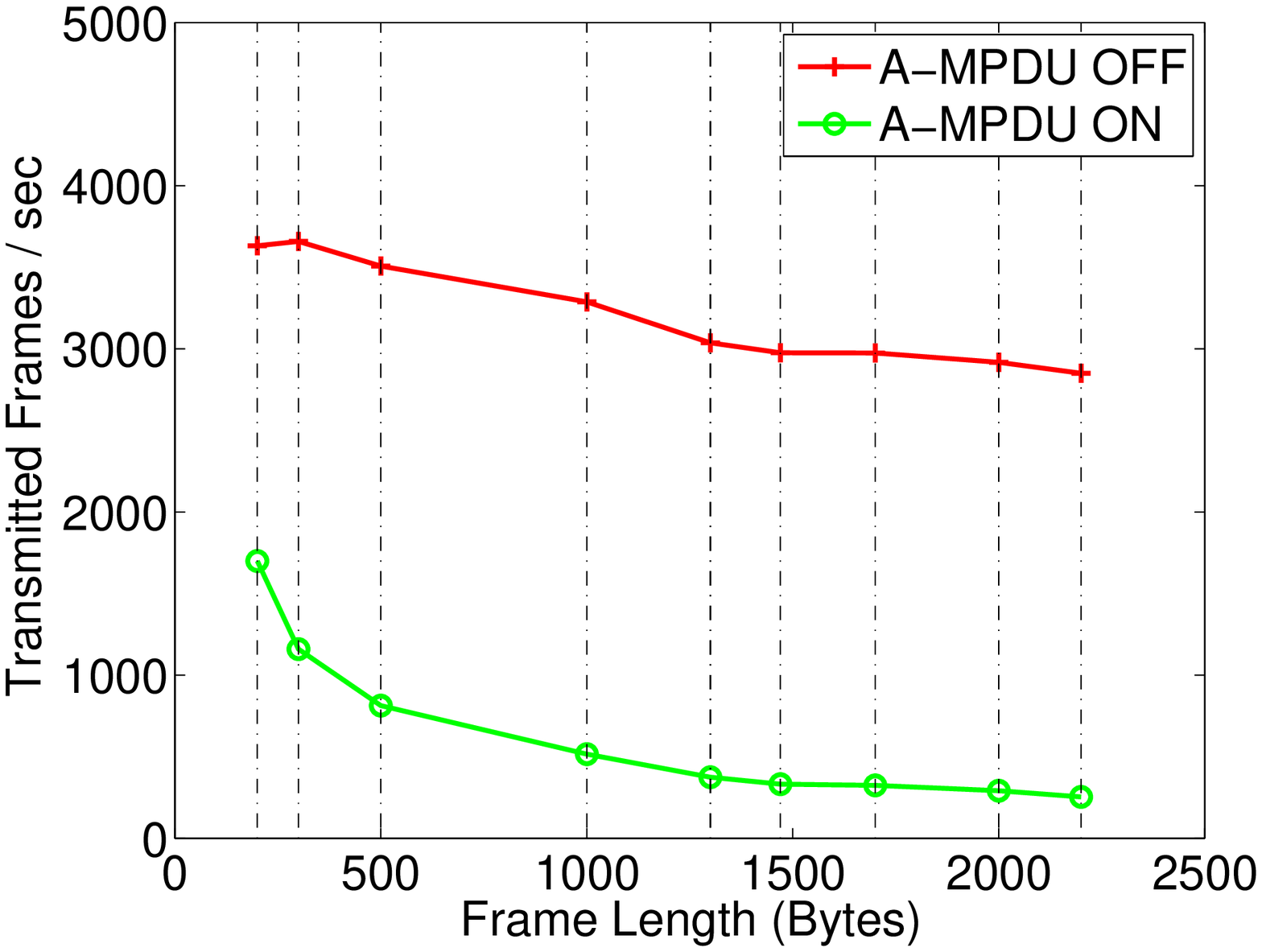}
      \label{fig: FRM_LEN_MIMO}}
      \hspace{0.22in}
                        \centering
            \subfigure[Atom Energy consumption/bit]{
                  \centering
      \includegraphics[width=0.6\columnwidth]{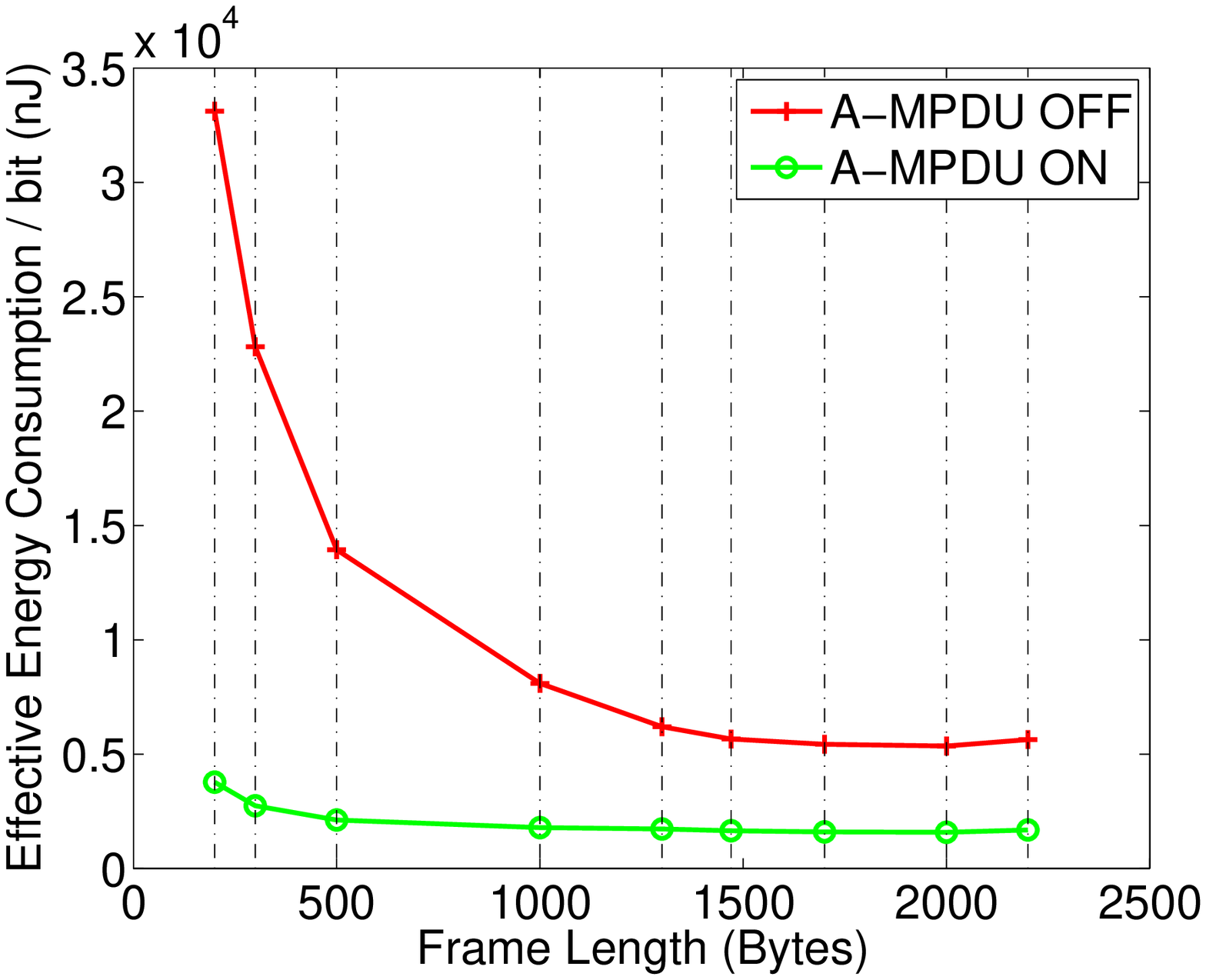}
      \label{fig: EB_LEN_MIMO}}
                   \vspace{-0.4cm}
          \caption{Performance of AR9380 NIC across varying Frame Length values}
           \vspace{-0.6cm}
      \end{figure*}  
       
Similar results are obtained while evaluating the impact of A-MPDU aggregation
on the consumption of the AR9380 NIC and the Atom node, which are plotted
in Fig. \ref{fig: AGG_NIC} and Fig. \ref{fig: AGG_ATOM} accordingly.
We clearly notice that the remarkably increased throughput performance that A-MPDU aggregation results in
for traffic loads above 100 Mbps does not come at higher energy costs, in comparison with the performance of 802.11a/g.
The Effective $E_{B}$ representation in Fig. \ref{fig: EB_AG}, summarises the above results and shows
that A-MPDU aggregation can increase energy efficiency up to 70$\%$.
Moreover, through comparison of power consumption at both the NIC and total node levels between the two standards,
we clearly notice that the supported by 802.11 high sampling rates, in accordance with low power consumption in idle mode, result in reduced average power consumption.
Direct comparison of $E_{B}$ values at the corresponding saturation points of each standard shows
that 802.11n offers more that 75$\%$ reduction of $E_{B}$ compared with the 802.11a/g standard.

\vspace{-0.1cm}
\subsubsection{Varying Frame Length}
\vspace{-0.1cm}      
Extensive throughput experiments were also conducted under varying frame lengths.
In order to enable delivery of frames longer than 1500 bytes to the MAC layer,
we configured the NIC's MTU size to the maximum supported value of 2274 bytes.
In this section, we refrain from presenting power consumption measurements as variation across varying lengths was minimal.
The throughput performance of AR5424 NIC is illustrated in  Fig. \ref{fig: THR_LEN_AG},
while Fig. \ref{fig: FRM_LEN_AG} and Fig. \ref{fig: EB_LEN_AG} represent the frame transmission rate and the Atom-node effective $E_{B}$ accordingly.
We observe that increasing frame length values result in improved throughput and energy efficiency performance.
However, in the case that \emph{FF} is enabled, aggregation of frames longer than 1700 bytes could not be handled by the driver,
as transmission duration exceeded the standard's 4 ms time threshold and thus no transmissions could be performed.

Throughput performance improvement and reduction of frame transmission rate across increasing length values
are also monitored when using the 802.11n NIC, as plotted in Fig. \ref{fig: THR_LEN_MIMO} and  Fig. \ref{fig: FRM_LEN_MIMO} accordingly.
Moreover, 802.11n protocol operation with enabled A-MPDU aggregation is able to use the highest frame lengths,
as its only limitations are the maximum number of subframes (64) and maximum A-MPDU length (65.535 bytes). 
Considering the results plotted in \ref{fig: THR_LEN_MIMO}, we observe that in the enabled A-MPDU case, even frames of 1000 bytes
are able to deliver high throughput performance (>130 Mbps), while longer frame lengths result in minimal energy efficiency improvement, as shown in in Fig. \ref{fig: EB_LEN_MIMO}.
Comparing the performance of the two protocols, we remark that 
802.11n with enabled A-MPDU is able to reduce energy expenditure by more than 70$\%$ for all considered frame lengths above 500 bytes,
when compared with the lowest achievable $E_{B}$ performance of 802.11a/g.
This observation yields interesting insights and motivates further investigation regarding the performance of A-MPDU aggregation under low quality channel conditions,
where low frame lengths are preferable due to the increased frame delivery rate.

\vspace{-0.1cm} 
\section{Conclusions and Future Work}
In this work, we presented detailed results that characterise how the evolution of the IEEE 802.11 standard
has impacted the energy efficiency of wireless devices.
Experimental evaluation considered the impact of several MAC-layer enhancements 
on the energy consumption of wireless transceivers and total nodes as well.
Our detailed findings can act as a benchmark for researchers pursuing energy efficient operation of wireless protocols.
As part of our future work, we plan on evaluating the energy efficiency of aggregation mechanisms under varying channel conditions
and complex network topologies.
       
\bibliographystyle{unsrt}
\small{
\bibliography{literature}
}
\end{document}